\documentclass[
  aps,
  prb,
  twocolumn,
  reprint,
  superscriptaddress,
  floatfix,
  citeautoscript,
  longbibliography,
]{revtex4-1}

\usepackage[version=4]{mhchem}
\usepackage{siunitx}
\usepackage{graphicx}
\usepackage[T1]{fontenc}
\usepackage[unicode,colorlinks=true,allcolors=blue]{hyperref}
\usepackage{cleveref}
\usepackage[para]{footmisc}
\usepackage{here}
\usepackage[dvipsnames,table]{xcolor}
\usepackage{amsmath}
\usepackage{mathtools}
\usepackage{amssymb}
\usepackage{bm}

\usepackage{cleveref}

\usepackage{microtype}

\usepackage{glossaries}
\glsdisablehyper
\newglossary[slg]{symbolslist}{syi}{syg}{Symbols} 

\glsaddkey{unit}{\glsentrytext{\glslabel}}{\glsentryunit}{\GLsentryunit}{\glsunit}{\Glsunit}{\GLSunit}

\newcommand{\gs}[1]{\glsunit{sym:#1}}

\newcommand{\newsymbolentry}[3]{\newglossaryentry{#1}{name=\ensuremath{#2}, description={#3}, type=symbolslist}}			
\newcommand{\newsymbolentryu}[4]{\newglossaryentry{#1}{name=\ensuremath{#2}, description={#3 [\si{#4}]}, type=symbolslist}}	

\newsymbolentry{sym:aB}{a_B}{Initial asymmetry of backward detector}
\newsymbolentry{sym:aF}{a_F}{Initial asymmetry of forward detector}
\newsymbolentry{sym:alpha}{\alpha}{Ratio of detector count rates without polarization}
\newsymbolentry{sym:asym4}{\mathcal{A}_4}{Asymmetry using four counter method}
\newsymbolentry{sym:asym_single}{\mathcal{A}}{Asymmetry using two counter method}

\newsymbolentry{sym:beta}{\beta}{Stretching exponent in stretched exponential (\Cref{eq:strexp}). }
\newsymbolentry{sym:beta_counter}{\beta}{Ratio of detector initial asymmetries without polarization}
\newsymbolentry{sym:Bcount}{\mathcal{B}}{Number of $e^-$ counted at the backward detector}
\newsymbolentryu{sym:Bfield}{\bm{B}}{External magnetic field}{\tesla}
\newsymbolentryu{sym:Beff}{\bm{B}_\mathrm{eff}}{Effective magnetic field due to \gs{B0} and \gs{B1} in the rotating reference frame (\Cref{eq:Beff})}{\tesla}
\newsymbolentryu{sym:B0}{\bm{B}_0}{External static magnetic field, assumed to be along \gs{zhat}}{\tesla}
\newsymbolentryu{sym:B0_sclr}{B_0}{Magnitude of static magnetic field}{\tesla}
\newsymbolentryu{sym:B1}{\bm{B}_1}{External RF oscillating magnetic field (\Cref{eq:B1_rotating})}{\tesla}
\newsymbolentryu{sym:B1_sclr}{B_1}{Magnitude of RF oscillating magnetic field (\Cref{eq:B1_rotating})}{\tesla}

\newsymbolentryu{sym:c}{c}{Speed of light}{\m/\s}

\newsymbolentry{sym:dasym}{\Delta \mathcal{A}}{Helicity-averaged two-counter asymmetry}
\newsymbolentry{sym:delta}{\delta}{Equal to $\frac{\gs{alpha}-1}{\gs{alpha}+1}$}
\newsymbolentryu{sym:Delta}{\Delta}{Beam on pulse duration}{\s}

\newsymbolentryu{sym:e}{e}{Electron charge}{C}
\newsymbolentry{sym:Ea}{E_A}{Arrhenius or \glstext{vft} activation energy}
\newsymbolentry{sym:efg_asym}{\eta}{Electric field gradient asymmetry (\Cref{eq:efg_asym})}
\newsymbolentryu{sym:efg_eq}{eq}{Electric field gradient strength}{\V\per\m^2}

\newsymbolentry{sym:Fcount}{\mathcal{F}}{Number of $e^-$ counted at the forward detector}

\newsymbolentryu{sym:gamma}{\gamma}{Gyromagnetic ratio}{\MHz/\tesla}

\newsymbolentry{sym:H0}{\mathcal{H}_0}{Hamiltonian related to coupling with a large static applied field, denoted by \gs{B0}.}
\newsymbolentry{sym:HQ}{\mathcal{H}_Q}{Hamiltonian related to quadrupolar coupling to the electric field gradient}

\newsymbolentry{sym:ihat}{\hat{\imath}}{Rotating reference frame coordinate}
\newsymbolentry{sym:I_mom}{I}{Dimensionless angular momentum $\bm{I}\equiv \bm{J}/\hbar$}
\newsymbolentry{sym:Iop}{I}{Spin operator}
\newsymbolentry{sym:I}{I}{Nuclear spin}

\newsymbolentry{sym:jhat}{\hat{\jmath}}{Rotating reference frame coordinate}
\newsymbolentryu{sym:tang_mom}{\bm{J}}{Total angular momentum}{\kg\cdot\m^2\per\s}
\newsymbolentry{sym:J}{\mathcal{J}}{Spectral density}

\newsymbolentry{sym:kB}{k_B}{Boltzmann constant}
\newsymbolentry{sym:khat}{\hat{k}}{Rotating reference frame coordinate}

\newsymbolentryu{sym:mag_mom}{\bm{\mu}}{Nuclear magnetic moment}{\A\m^2}

\newsymbolentryu{sym:nu0}{\nu_0}{Larmor frequency (\Cref{eq:larmor})}{\Hz}
\newsymbolentryu{sym:nuQ}{\nu_Q}{Quadrupole splitting frequency (\Cref{eq:nuQ})}{\Hz}
\newsymbolentry{sym:Nt}{N(t)}{The total number of \eli\ in the sample at time $t > t'$}

\newsymbolentry{sym:P}{\mathcal{P}}{Polarization at time $t$ averaged over all implantation times $t'$}

\newsymbolentryu{sym:Q}{Q}{Nuclear quadrupole moment. Often quoted in units of barns, where a factor of \gs{e} is dropped. The quantity $eQ$ has the proper units}{Cm^2}

\newsymbolentryu{sym:R0}{R_0}{Implantation rate into sample}{\per\s}
\newsymbolentryu{sym:RF}{R_F}{Implantation rate into forward detector}{\per\s}
\newsymbolentryu{sym:RB}{R_B}{Implantation rate into backward detector}{\per\s}

\newsymbolentry{sym:stern}{\gamma_\infty}{Sternheimer antishielding factor (\Cref{eq:stern})}

\newsymbolentryu{sym:tprime}{t'}{Nuclear implantation time}{\s}
\newsymbolentryu{sym:tau}{\tau}{Nuclear lifetime}{\s}
\newsymbolentryu{sym:tauc}{\tau_c}{Microscopic molecular correlation time}{\s}
\newsymbolentryu{sym:tauprime}{\tau'}{Equal to $(1/\gs{tau}+1/\gs{T1})^{-1}$}{\s}
\newsymbolentryu{sym:T1}{T_1}{Exponential spin-lattice relaxation time of longitudinal magnetization:  \Cref{eq:gp:bloch}}{\s}
\newsymbolentryu{sym:T2}{T_2}{Exponential spin-spin relaxation time of transverse magnetization: \Cref{eq:gp:bloch}}{\s}
\newsymbolentryu{sym:Tg}{T_\mathrm{g}}{Glass transition temperature}{\K}
\newsymbolentryu{sym:Tm}{T_\mathrm{m}}{Melting temperature}{\K}
\newsymbolentryu{sym:Tk}{T_\mathrm{k}}{Kauzmann temperature}{\K}
\newsymbolentryu{sym:Tvft}{T_\mathrm{VFT}}{\glsdesc{vft} temperature}{\K}

\newsymbolentryu{sym:v}{v}{Velocity of emitted $e^-$ in nuclear decay}{\m/\s}
\newsymbolentryu{sym:efg}{V}{Electric field gradient component (\Cref{eq:efg_elements})}{\V/\m^2}

\newsymbolentryu{sym:w0}{\omega_0}{Larmor frequency}{rad\per\s}
\newsymbolentry{sym:wt}{W(\theta)}{Probability density of beta emission angle $\theta$}
    
\newsymbolentry{sym:xhat}{\hat{x}}{Lab frame coordinate}

\newsymbolentry{sym:yhat}{\hat{y}}{Lab frame coordinate}

\newsymbolentry{sym:zhat}{\hat{z}}{Lab frame coordinate}   

\newacronym{ac}{AC}{alternating current}
\newacronym{afm}{AFM}{atomic force microscopy}
\newacronym{alc}{ALC}{avoided level crossing}
\newacronym{api}{API}{application programming interface}
\newacronym{ariel}{ARIEL}{Advanced Rare Isotope Laboratory}
\newacronym{arpes}{ARPES}{angle-resolved photoemission spectroscopy}

\newacronym[sort={b-NMR}]{bnmr}{$\beta$-NMR}{$\beta$-detected \glstext{nmr}}
\newacronym[sort={b-NQR}]{bnqr}{$\beta$-NQR}{$\beta$-detected nuclear quadrupole resonance}
\newacronym{bca}{BCA}{binary collision approximation}
\newacronym{bpp}{BPP}{Bloembergen-Purcell-Pound}
\newacronym{bsc}{BSC}{\ce{Bi2Se3:Ca}}
\newacronym{btm}{BTM}{\ce{Bi2Te3:Mn}}
\newacronym{bts}{BTS}{\ce{Bi2Te2Se}}

\newacronym{camp}{CAMP}{control and monitor program}
\newacronym{ccd}{CCD}{charge-coupled device}
\newacronym{cdw}{CDW}{charge density wave}
\newacronym{cgs}{CGS}{centimetre-gram-second system of units}
\newacronym{cmms}{CMMS}{Centre for Molecular and Materials Science}
\newacronym{codata}{CODATA}{Committee on Data for Science and Technology}
\newacronym{cpu}{CPU}{central processing unit}
\newacronym{create}{CREATE}{Collaborative Research and Training Experience Program}
\newacronym{cw}{CW}{continuous wave}

\newacronym{daq}{DAQ}{data acquisition}
\newacronym{dc}{DC}{direct current}
\newacronym{dft}{DFT}{density functional theory}
\newacronym{dos}{DOS}{density of states}
\newacronym{dqt}{DQT}{double-quantum transition}

\newacronym{efg}{EFG}{electric field gradient}
\newacronym{epr}{EPR}{electron paramagnetic resonance}
\newacronym{esr}{EPR}{electron spin resonance}
\newacronym{endor}{ENDOR}{electron nuclear double resonance}
\newacronym{emim-ac}{\ce{EMIM-Ac}}{1-ethyl-3-methylimidazolium acetate}
\newacronym{epics}{EPICS}{Experimental Physics and Industrial Control System}

\newacronym{fft}{FFT}{fast Fourier transform}
\newacronym{fom}{FoM}{figure of merit}
\newacronym{fwhm}{FWHM}{full width at half maximum}

\newacronym{gga}{GGA}{generalized gradient approximation}
\newacronym{gui}{GUI}{graphical user interface}

\newacronym{hv}{HV}{high-voltage}
\newacronym{hwhm}{HWHM}{half width at half maximum}

\newacronym{is}{IS}{impedance spectroscopy}
\newacronym{isac}{ISAC}{isotope separator and accelerator}
\newacronym{isol}{ISOL}{isotope separation online}
\newacronym{isosim}{IsoSiM}{Isotopes for Science and Medicine}

\newacronym{kww}{KWW}{Kohlrausch-Williams-Watts}

\newacronym{lcao}{LCAO}{linear combination of atomic orbitals}
\newacronym{lda}{LDA}{local density approximation}
\newacronym{leis}{LEIS}{low-energy ion scattering}
\newacronym{lib}{LIB}{lithium-ion battery}
\newacronym{lsat}{LSAT}{\ce{(La,Sr)(Al,Ta)O3}}

\newacronym{mas}{MAS}{magic angle spinning}
\newacronym{mbe}{MBE}{molecular beam epitaxy}
\newacronym{md}{MD}{molecular dynamics}
\newacronym{midas}{MIDAS}{Maximum Integrated Data Acquisition System}
\newacronym{mit}{MIT}{metal-insulator transition}
\newacronym{mnr}{MNR}{Meyer-Neldel rule}
\newacronym{mqt}{mqt}{multi-quantum transition}
\newacronym{mud}{MUD}{Muon Data}

\newacronym{nbm}{NBM}{neutral beam monitor}
\newacronym{neb}{NEB}{nudged elastic band}
\newacronym{nim}{NIM}{nuclear instrumentation module}
\newacronym{nmr}{NMR}{nuclear magnetic resonance}
\newacronym{no}{NO}{nuclear orientation}
\newacronym{nqr}{NQR}{nuclear quadrupole resonance}
\newacronym{nserc}{NSERC}{Natural Sciences and Engineering Research Council of Canada}

\newacronym{oa}{OA}{optical absorption}
\newacronym{ode}{ODE}{ordinary differential equation}

\newacronym{pac}{PAC}{perturbed angular correlation}
\newacronym{pad}{PAD}{perturbed angular distribution}
\newacronym{pld}{PLD}{pulsed laser deposition}
\newacronym{ps}{PS}{polystyrene}
\newacronym{pypi}{PyPI}{Python Package Index}

\newacronym{qens}{QENS}{quasielastic neutron scattering}
\newacronym{ql}{QL}{quintuple layer}
\newacronym{qo}{QO}{quantum oscillations}

\newacronym{rbs}{RBS}{Rutherford backscattering}
\newacronym{rf}{RF}{radio frequency}
\newacronym{rheed}{RHEED}{reflection high-energy electron diffraction}
\newacronym{rib}{RIB}{radioactive ion beam}
\newacronym{rtil}{RTIL}{room temperature ionic liquid}

\newacronym{sae}{SAE}{spin-alignment echo}
\newacronym{si}{SI}{International System of Units}
\newacronym{sims}{SIMS}{secondary ion mass spectrometry}
\newacronym{slr}{SLR}{spin-lattice relaxation}
\newacronym{snr}{\textit{S}/\textit{N}}{signal-to-noise ratio}
\newacronym{squid}{SQUID}{superconducting quantum interference device}
\newacronym{srim}{SRIM}{Stopping and Range of Ions in Matter}
\newacronym{ssid}{SSID}{solid-state ionic device}
\newacronym{ssr}{SSR}{spin-spin relaxation}
\newacronym{stm}{STM}{scanning tunnelling microscopy}
\newacronym{sts}{STS}{scanning tunnelling spectroscopy}

\newacronym{ti}{TI}{topological insulator}
\newacronym{trim}{TRIM}{Transport and Range of Ions in Matter}
\newacronym{tss}{TSS}{topological surface state}
\newacronym{tmd}{TMD}{transition metal dichalcogenide}

\newacronym{uhv}{UHV}{ultra-high vacuum}

\newacronym{vdw}{vdW}{van der Waals}
\newacronym{vft}{VFT}{Vogel-Fulcher-Tammann}

\newacronym{xrd}{XRD}{x-ray diffraction}
\newacronym{xrr}{XRR}{x-ray reflection}

\newacronym{ybco}{YBCO}{\ce{YBa2Cu3O_{6+x}}}
\newacronym{ysz}{YSZ}{yttria-stabilized zirconia}

\newacronym[sort={muSR}]{musr}{$\mu$SR}{muon spin rotation}
\newacronym{alc-musr}{ALC-μSR}{avoided level crossing muon spin rotation}
\newacronym{le-musr}{LE-μSR}{low energy muon spin rotation}
\newacronym{lf-musr}{LF-μSR}{longitudinal field muon spin rotation}
\newacronym{rf-musr}{RF-μSR}{radio frequency muon spin rotation}
\newacronym{tf-musr}{TF-μSR}{transverse field muon spin rotation}
\newacronym{zf-musr}{ZF-μSR}{zero field muon spin rotation}

\usepackage[inline,final]{showlabels}

\usepackage{comment}

\usepackage{listings}
\usepackage{color}
\usepackage[dvipsnames]{xcolor}
\usepackage{textcomp}
\usepackage{algcompatible}
\usepackage[linesnumbered,ruled,vlined]{algorithm2e} 

\begin{document} 

\title{Digging Into MUD With Python: \\mudpy, bdata, and bfit}

\newcommand{\bnmr}{\gls{bnmr}}
\newcommand{\bnqr}{\gls{bnqr}}
\newcommand{\musr}{\gls{musr}}
\newcommand{\elip}{$^8$Li$^+$}
\newcommand{\eli}{$^8$Li}
\newcommand{\lip}{Li$^+$}

\newcommand{\ubcsbqmi}{Stewart Blusson Quantum Matter Institute, University of British Columbia, Vancouver, BC V6T~1Z4, Canada}
\newcommand{\ubcchem}{Department of Chemistry, University of British Columbia, Vancouver, BC V6T~1Z1, Canada}
\newcommand{\ubcphas}{Department of Physics and Astronomy, University of British Columbia, Vancouver, BC V6T~1Z1, Canada}
\newcommand{\triumf}{TRIUMF, 4004 Wesbrook Mall, Vancouver, BC V6T~2A3, Canada}

\newcommand\mycommfont[1]{\small\ttfamily\textcolor{Red}{#1}}
\SetCommentSty{mycommfont}

\author{Derek~Fujimoto}
\email[]{fujimoto@phas.ubc.ca}
\affiliation{\ubcsbqmi}
\affiliation{\ubcphas}

\date{\today}

\begin{abstract}
Used to store the results of \musr\ measurements at TRIUMF, the \gls{mud} file format serves as a useful and flexible scheme that is both lightweight and self-describing. The \gls{api} for these files is written in C and FORTRAN, languages not known for their ease of use. In contrast, Python is a language which emphasizes rapid prototyping and readability. This work describes three Python 3 packages to interface with \gls{mud} files and analyze their contents: mudpy, bdata, and bfit. The first enables easy access to the contents of any \gls{mud} file. The latter two are implemented specifically for the implanted-ion \bnmr\ experiment at TRIUMF. These tools provide both an \gls{api} and \gls{gui} to help users extract and fit \bnmr\ data.
\end{abstract}

\maketitle
\glsresetall


\section{Motivation}

The first \musr\ measurements were recorded in 1957, at the Nevis cyclotron in the United States of America.\cite{Garwin1957a,Garwin2003} While the field has thrived over its long history, the technique remains restricted to large nationally-supported facilities.\cite{Brewer2012} Today, there are only a handful of locations capable of producing the particle beam needed to conduct \musr, including: TRIUMF, Canada; ISIS, located in the United Kingdom; PSI in Switzerland; and the Japanese facility J-PARC. The \gls{mud} file format is used to store \musr\ data taken at TRIUMF.\cite{Whidden1994} This is a self-describing binary format (i.e. not ASCII), containing the measurement data, device settings, experimental conditions such as the temperature or the magnetic field, and some metadata. 

As with many older science applications, the \gls{mud} file \gls{api} is written in C and FORTRAN. These statically-typed and compiled languages are known for their computational efficiency, but can be difficult to work with. This is perhaps one of the reasons why scientific computing has, in many communities, shifted to more modern languages such as Python: a dynamically-typed and interpreted language. As a result, Python has amassed a massive library of data analysis tools.\cite{Virtanen2020,McKinney2010,scikit-learn} The primary advantage of Python is the short development time of programs written in the language. This is particularly important in the context of scientific analysis, which are typically run only a few times by select individuals. As a result, the time taken to write the analysis code is a large part of the program's effective run time. The aim of this work is to bring this rapid prototyping style of analysis to the \musr\ and \bnmr\ communities. 

It should be acknowledged that a large body of analysis software exists to support \musr\ workers. Examples include WIMDA,\cite{Pratt2000} an older Windows application; MANTID,\cite{Arnold2014} developed by and for ISIS; and Musrfit,\cite{Suter2012} maintained by the workers at PSI. Data stored in the \gls{mud} format are compatible with Musrfit. These programs are quite powerful,\cite{Locans2018} but can be cumbersome outside of their intended scope (e.g. when developing new methods\cite{Simoes2020}). The packages introduced here are very lightweight, providing a simple interface to any other Python package, allowing for a great deal of flexibility and sophistication. Like many Python packages, those described in this work are freely distributed through the \gls{pypi} and GitHub.\footnote{The packages are listed on \glstext{pypi} (\url{https://pypi.org/}) as \texttt{mud-py}, \texttt{bdata}, and \texttt{bfit}.} This trivializes installation and maintenance by installing missing dependencies, updating packages, and providing a consistent method of version tracking. This is in stark contrast to another popularly used framework, ROOT,\cite{Brun1997a} which serves as the basis for Musrfit, and whose set up process can be quite involved. 

A closely related technique to \musr, \bnmr\ relies on the same physics principles but uses a radioactive isotope, rather than a muon. At present, the only active and permanent implanted-ion \bnmr\ spectrometer is at TRIUMF.\cite{Morris2014a,MacFarlane2015,Kiefl2018} Unlike \musr, \bnmr\ does not have an extensive suite of analysis programs well-suited to the specifics of the technique, however it still uses the \gls{mud} format as the basis of its \gls{daq} system. While there have been some recent improvements to this situation,\cite{Saadaoui2018} the analysis required for any non-trivial \bnmr\ experiment necessitates the development of new code to meet the individual requirements of each experiment.\cite{Szunyogh2018,Fujimoto2019,McFadden2017a,Karner2019} The packages described in this work immensely expedite this process.

\section{Software Description}

This work describes three Python 3 packages: mudpy, bdata, and bfit. The former is a general-purpose \gls{mud} file reader, whereas the latter two are specific to the \bnmr\ experiment at TRIUMF.\cite{Morris2014a,MacFarlane2015} Both mudpy and bdata serve to contribute an object layer between the file and analysis code, whereas bfit implements fitting functionality. 

The mudpy package provides a wrapper for each of the C functions in the \texttt{mud\_friendly}\cite{Whidden1994} \gls{api} using Cython,\cite{Behnel2011} and a Python class, \texttt{mdata}, which automates file access, saving the contents as object attributes. \Gls{mud} files store data of five different types: description, histogram, independent variable, scaler, and comment. The description type contains the file metadata and is saved directly as \texttt{mdata} attributes. The remainders are saved as containers organized into specialized dictionaries, as demonstrated in the example section. The histograms contain the primary data: counts from the various detectors needed to measure the nuclear spin polarization. The independent variables may contain experiment settings or measurements such as the temperature. The scalers contain information from secondary readouts of the detectors: total number of counts and the most recent reading. Comments are additional notes from the experimenters. 

The \texttt{bdata} object inherits from \texttt{mdata}, providing additional functionality for variable lookup, calculation of the beta-decay asymmetry, and remote fetching of data from the archive.\footnote{All non-commercial data taken at TRIUMF can be found at \url{http://cmms.triumf.ca/}.} The bdata package also provides the classes \texttt{bjoined} and \texttt{bmerged} for concatenating and merging \texttt{bdata} objects, respectively. 

The bfit package provides a suite of analysis tools for \bnmr\ experiments, accessible through both the Python \gls{api} and a \gls{gui}. The primary goal of the \gls{gui} is to enable inexperienced programmers and external users to do \bnmr, however, it also makes simple analyses very fast and convenient, useful even for experienced users. The \gls{gui} supports a number of features, such as: shared parameter fitting, interactive and graphical selection of initial fit parameters, assigning functional constraints to fit parameters (e.g. if a variable is a function of temperature, as is the case for Korringa relaxation\cite{Korringa1950}), periodic fetching and redrawing of data, and displaying the file contents in a format easily compatible with the MIDAS \gls{daq} system.\cite{Ritt1993} bfit also supports the dynamic importing of user-defined $\chi^2$ minimizers, allowing for the use of other codes, such as ROOT, in fitting data. The default $\chi^2$ minimizer in bfit is a Levenberg-Marquardt algorithm\cite{More1978} when no bounds are set, or if the parameter space is bounded, then a Trust Region Reflective algorithm is used.\cite{Branch1999} Both are implemented using the \texttt{scipy.optimize.curve\_fit} function.\cite{Virtanen2020} The graphical interface for bfit is provided by TkInter, and plotting functionality is implemented using Matplotlib\cite{Hunter2007}. Numerical calculations employ the use of SciPy,\cite{Virtanen2020} NumPy,\cite{VanderWalt2011a} and Pandas.\cite{McKinney2010} These packages run C or FORTRAN code under the hood, providing a huge reduction in computation time. 

Given that mudpy exists only to read, write, and store the contents of files, the remainder of this work will focus on the \bnmr-specific implementations. Of critical importance is the calculation of the asymmetry of the average beta-decay direction. This discussion relates the measured asymmetry to the underlying polarization and shows how the fit functions are constructed in the case of pulsed beam measurements. Next, an overview of the global fitting method is presented. Lastly, a few illustrative examples using the three packages will be given. 

\section{$\beta$-NMR Polarization and Asymmetry}

Similar to \musr, implanted-ion \bnmr\ measures the nuclear spin polarization of an implanted ensemble of radioactive particles through their anisotropic beta decay.\cite{MacFarlane2015} A major distinction, however, is that the lifetime ($\tau$) of the isotopes used in \bnmr\ are much longer than that of the muon. For example, \eli, the typical \bnmr\ probe at TRIUMF, has $\tau=\SI{1.21}{\s}$, as compared to $\tau=\SI{2.2}{\micro\s}$ for the muon. The coarser time resolution of \bnmr\ results in a very different analysis, despite the similarities in the physical principles. 

The probability that an $e^-$ is emitted at angle $\theta$, relative to the spin of the probe nucleus, is 
\begin{equation}
	\label{eq:Wtheta}
\gs{wt} = 1+\frac{\gs{v}}{\gs{c}}PA\cos(\theta),
\end{equation}
where $P$ is the polarization of the nuclear ensemble, $v$ is the speed of the emitted $e^-$, $c$ is the speed of light, and $|A|<1$ is an asymmetry parameter.\cite{Jackson1957,Correll1983} For \eli, $A=-\tfrac{1}{3}$\cite{Arnold1988} and the in-flight nuclear spin polarization along the beam axis at TRIUMF is approximately \SI{70}{\%}.\cite{Levy2002} The ions are implanted at an approximately constant rate, \gs{R0}, which is switched on at $t=0$. Let $N(t,\gs{tprime})dt'$ be the number of nuclei which arrived in the interval (\gs{tprime}, $\gs{tprime} + dt'$), and survive until time $t$:
\begin{equation} \label{eq:N}
N(t,\gs{tprime})dt' = R_0\exp[-(t-t')/\tau]dt',
\end{equation}
where \gs{tau} is the nuclear lifetime.\cite{KieflNotes} The total number of nuclei in the sample at time $t>\gs{tprime}$ is then
\begin{equation} \label{eq:pol_norm}
\begin{split}
\gs{Nt} &= \int_0^tN(t,\gs{tprime})dt'\\
&= \gs{R0}\gs{tau} [1-\exp(-t/\gs{tau})].
\end{split}
\end{equation}
Let $p(t,\gs{tprime})$ be the average polarization of an ensemble of probes implanted at $t'$, at the moment of decay at time $t>\gs{tprime}$. The simplest case is when $p(t,\gs{tprime})$ is exponential, $\exp[(t-t')/T_1]$, however in general it may be any function of $t$ and \gs{tprime}. Accounting for all arrival times, the average polarization at time $t>\gs{tprime}$ is\cite{KieflNotes}
\begin{equation}\label{eq:pol}
\begin{split}
\gs{P}(t) &= \frac{1}{\gs{Nt}}\int_0^t N(t,\gs{tprime})p(t,t')dt'\\
&= \frac{\int_0^t\exp[-(t-\gs{tprime})/\gs{tau} ]p(t,\gs{tprime})dt'}{\gs{tau} [1-\exp(-t/\gs{tau})]}.
\end{split}
\end{equation}
The polarization of the implanted ensemble is measured by counting the emitted betas in the forward ($F$) and backward ($B$) directions, relative to the beam direction. If $f(t,\gs{tprime})dt'$ is the number of betas detected in detector $F$ during the time interval ($t$, $t+dt$), then
\begin{equation}
f(t,\gs{tprime})dt' = \tfrac{1}{2}N(t,\gs{tprime})[1+ap(t,\gs{tprime})]dt',
\end{equation}
where $a$ is a constant of proportionality.\cite{KieflNotes} The average number of betas arriving in detector $F$ is then: 
\begin{equation}
\begin{split}		
\gs{Fcount}(t) &= \int_0^t f(t,\gs{tprime})dt'\\
& = \frac{\gs{R0}}{2} \gs{tau} [1-\exp(-t/\gs{tau})](1+a\gs{P}(t)),
\end{split}
\end{equation}
and similarly for $\gs{Bcount}(t)$, where $a\rightarrow -a$. In principle, differences in the detectors may lead to unique values of $R_0$ and $|a|$ for each. However, if they are the same, then the asymmetry is proportional to the polarization:
\begin{equation}\label{eq:asym} 
\gs{asym_single}(t) \equiv \frac{\gs{Fcount}(t)-\gs{Bcount}(t)}{\gs{Fcount}(t)+\gs{Bcount}(t)}
= a\gs{P}(t).
\end{equation}
If the rates, $R_0$, are not the same for each detector, then
\begin{equation}
\gs{asym_single}(t) = \frac{\gs{delta}+a\gs{P}(t)}{1+\gs{delta} a\gs{P}(t)},
\end{equation}
where $\gs{alpha} \equiv \frac{\gs{RF}}{\gs{RB}}$ and $\gs{delta}=\frac{\gs{alpha}-1}{\gs{alpha}+1}$.\cite{KieflNotes} The effects produced by the non-ideal case of $\alpha\neq1$ are typically reduced by combining the asymmetry with that of the inverted spin polarization state (denoted by $\pm$):\cite{KieflNotes}
\begin{equation}  \label{eq:asym_diff}
\gs{dasym}(t) = \frac{\gs{asym_single}^+(t)-\gs{asym_single}^-(t)}{2}
= \frac{a\gs{P}(t)(1-\gs{delta}^2)}{1-[\gs{delta} a\gs{P}(t)]^2}.
\end{equation}
Since polarization inversion is equivalent to $\mathcal{P}_+=-\mathcal{P}_-$, and in the absence of spectrometer-related distortions $a_F = -a_B$, then $\gs{Fcount}^\pm=\gs{Bcount}^\mp$. Physically, if the spins are pointed at detector $F$ in the ($+$) state, then in the ($-$) state they are pointed at detector $B$, so the signal is invariant if both polarization and detectors are exchanged. Therefore, one can take the geometric means of these pairings to form a 4-counter asymmetry:\cite{Widdra1995}
\begin{equation}\label{eq:asym_4counter} 
	\gs{asym4}(t) = \frac{\sqrt{\gs{Fcount}^+\gs{Bcount}^-}-\sqrt{\gs{Fcount}^-\gs{Bcount}^+}}{\sqrt{\gs{Fcount}^+\gs{Bcount}^-}+\sqrt{\gs{Fcount}^-\gs{Bcount}^+}}.
\end{equation}
The advantage of this formulation is that the terms $\sqrt{\gs{Fcount}^\pm\gs{Bcount}^\mp}$ share the coefficient $\sqrt{\gs{RF}\gs{RB}}$, which cancels and eliminates any dependence on the rate. The 4-counter asymmetry is proportional to the polarization with the same scaling factor as \Cref{eq:asym}, regardless of $\alpha$. 

We now consider the case where the scaling factors of the two detectors also differ: $\gs{beta_counter} \equiv \frac{\gs{aF} }{\gs{aB} } \neq 1$. In this case, 
\begin{equation} \label{eq:asym_diff_beta}
\gs{dasym}(t) = \delta_+\gs{aB} \gs{P}(t)\frac{1-\frac{\delta_-}{\delta_+}\Big(\frac{\gs{alpha}-1}{\gs{alpha}+1}\Big)}{1-\Big(\delta_-\gs{aB} \gs{P}(t)\Big)^2}
\end{equation}
where $\delta_\pm \equiv \frac{\gs{alpha}\gs{beta_counter}\pm1}{\gs{alpha}+1}$. In the limit where $\alpha\rightarrow0$ and $\beta\rightarrow0$,
\begin{equation}\label{eq:asym_diff_beta_lim} 
\begin{split}
\gs{dasym}(t) \approx~
&\frac{1}{2}\Big[(\gs{beta_counter}+1) - (\gs{alpha}-1)^2\Big]\gs{aB}\gs{P}(t) +\\
&\frac{1}{2}\Big[(\gs{beta_counter}-1) + (\gs{alpha}-1)\Big]^2\Big(\gs{aB}\gs{P}(t)\Big)^3.
\end{split}
\end{equation}
In contrast, if this generalization of $\gs{beta_counter}\neq 1$ is accounted for in \Cref{eq:asym_4counter}, we find that the dependence on $\gs{alpha}$ again is exactly cancelled, however the dependence on \gs{beta_counter} remains. To first order this is
\begin{equation}\label{eq:asym_4counter_beta}
\gs{asym4}(t) \approx \left(\frac{\gs{beta_counter}+1}{2}\right)\gs{aB}\gs{P}(t),
\end{equation}
which is the same as the first term of \Cref{eq:asym_diff_beta_lim} if $\gs{alpha}=1$. Differences in rate and scaling are also common sources of systematic error in \musr\ experiments.\cite{Riseman1994,Brewer1987}

The implementation of these calculations can be found in both bdata and bfit. In bdata, the asymmetry of each of the individual polarization states is calculated using \Cref{eq:asym}, whereas the combined asymmetry is calculated with \Cref{eq:asym_4counter}. The exponential and multi-exponential \gls{slr} fitting functions used in bfit are calculable analytically using \Cref{eq:pol}, whereas the stretched exponential relaxation fit function is computed numerically using a double-exponential integration scheme, important for its speed and stability at $t=0$.\cite{Cook2008,Mori2001}  Note that in order to measure the \gls{slr}, a pulsed beam is used; thus, when the beam is turned off at time $\Delta$, the integration limit of \Cref{eq:pol,eq:pol_norm} is set to $\Delta$ when $t>\Delta$.

\section{Global Fitting}

Global fitting refers to the procedure by which the parameters of best fit are found for a set of data, in the case where one or more of these parameters are shared across the data set. For example: the data set may consist of many Gaussian distributions which are known to be collectively described by a single mean, but have different standard deviations. The global fitting class defined in \texttt{bfit.fitting.global\_fitter} takes as input the data, the fitting function, and a list of boolean values specifying which parameters are to be shared. The global $\chi^2$, resulting from a simultaneous fit to all the distributions, is then minimized. The global fitting object also allows for the initial parameter values and bounds to be set intelligently, inferring from context whether the input refers to the parameter for a given data set, or should be applied to all data sets. 

Continuing with this example, if there are $N_\mathrm{dat}$ points in each distribution, and $N_\mathrm{set}$ distributions in the data set, then the data inputs to the constructor will each be 2D array-like objects with the shape $N_\mathrm{set}\times N_\mathrm{dat}$. The input for the function may be a single function handle or a list of handles of length $N_\mathrm{set}$, with the restriction that all functions must take the same inputs in the same order. The length of the list of booleans indicating sharing defines the number of parameters to be minimized. The remaining function inputs should be passed as constant values through the \texttt{metadata} parameter. The constructor may also take a 2D list of booleans indicating which parameters for which distribution are to be fixed to their initial value.  These latter two inputs are optional.


Like all the minimization in bfit, the global fitting utilizes the \texttt{scipy.optimize.curve\_fit} function, which has the basic prototype \texttt{curve\_fit(f, xdata, ydata)}. \texttt{f} is the fitting function of the type \texttt{f(x,...)}. The strategy is to define \texttt{f} such that it has access to the data through the object attributes, and inflate the 1D array of input parameters from \texttt{curve\_fit} to match. The call to \texttt{curve\_fit} is applied to the concatenation of the data arrays. This is described in detail by Algorithms~\ref{alg:getmap} and \ref{alg:master_eq}. A similar flattening procedure is applied to the initial parameters and fitting bounds, such that they can be passed to \texttt{curve\_fit}, and the inverse procedure is applied to inflate the results back to the original input shape. 

\begin{algorithm}
\SetAlgoLined
\DontPrintSemicolon
\SetKwComment{Comment}{\#}{}
\SetKwInOut{Input}{Input}
\SetKwInOut{Output}{Output}
\KwData{ \parbox[t]{0.9\columnwidth}{\raggedright
    Which indices are fixed and which are shared,\\
    The number of data sets, $N_\mathrm{set}$,\\
    The number of fit parameters, $N_\mathrm{par}$}}
\Input{None}
\Output{Matrix $L$ of indices, mapping a 1D arrangement to 2D}

\SetKwFunction{getmap}{get\_map}
\SetKwProg{Fn}{Function}{:}{}
\Fn{\getmap{void}}{

\BlankLine
$L = N_\mathrm{set}\times N_\mathrm{par}$ matrix\;
$\bm{u}=$ empty vector\;

\BlankLine
\For{$i \in \{0, 1, ..., N_\mathrm{set}-1\}$}
{
    \For{$j \in \{0, 1, ..., N_\mathrm{par}-1\}$}
    {
        \If{parameter j is shared}
        {
            $L_{ij} = j$\;
        }

        \BlankLine
        \ElseIf{parameter j is fixed}
        {
            $L_{ij} = -1-j - (i\cdot N_\mathrm{par})$\;
        }

        \BlankLine        
        \Else
        {
            $L_{ij} = j + (i\cdot N_\mathrm{par})$\;
        }

        \BlankLine        
        \If{$L_{ij}\ge0$ and $L_{ij}\notin \bm{u}$}
        {
            append $L_{ij}$ to $\bm{u}$\;
        }
        
    }
}
\BlankLine
\tcc{Remap indices to those from curve\_fit}
$L_{ij} = k$ if $L_{ij} = u_k~\forall~i,j,k$\;
$L_{ij} = N_\mathrm{set} \cdot N_\mathrm{par}-L_{ij}$ if $L_{ij} < 0 ~\forall~i,j$\;

\Return{L}\;
}
\textbf{end}\;
\caption{Get a matrix of indices mapping the 1D input to \texttt{curve\_fit} to a 2D arrangement}
\label{alg:getmap}
\end{algorithm}

\begin{algorithm}
\SetAlgoLined
\DontPrintSemicolon
\SetKwComment{Comment}{\#}{}
\SetKwInOut{Input}{Input}
\SetKwInOut{Output}{Output}
\KwData{
    \parbox[t]{0.9\columnwidth}{\raggedright
    $X$: $N_\mathrm{set}\times N_\mathrm{dat}$ matrix, independent variable\\
    $M$: $N_\mathrm{set}\times N_\mathrm{meta}$ matrix, additional inputs\\
    $P$: $N_\mathrm{set}\times N_\mathrm{par}$ matrix, initial parameters\\
    $f(\bm{x}, p_1, p_2, ..., p_{(N_\mathrm{par}+N_\mathrm{meta})})$: fitting function}}

\Input{From \texttt{curve\_fit}, vectors of values for the independent variable $\bm{x}$ (unused), and parameters $\bm{p}$ (length $N_\mathrm{par}$).}
\Output{An array of values corresponding to the concatenated \texttt{y} values}

\SetKwFunction{Master}{f}
\SetKwProg{Fn}{Function}{:}{}
\Fn{\Master{$\bm{x}$,$\bm{p}$}}{

$L =$ \getmap{}\;

\BlankLine
\tcc{Add initial parameters to input array}
\For{$i \in \{N_\mathrm{set}-1, N_\mathrm{set}-2, ..., 0\}$}
{
    \For{$j \in \{N_\mathrm{par}-1, N_\mathrm{par}-2, ..., 0\}$}
    {
        append $P_{ij}$ to $\bm{p}$\;
    }
}

\tcc{Get data for output}
Let $Y'$ be a $N_\mathrm{set}\times N_\mathrm{dat}$ matrix\;
\For{$i \in \{0, 1, ..., N_\mathrm{set}-1\}$}
{
    \For{$j \in \{0, 1, ..., N_\mathrm{dat}-1\}$}
    {   
        $Y'_{ij} = f($\parbox[t]{0.4\columnwidth}{\raggedright$X_{ij},$\\
        $p_{L_{(i,1)}},p_{L_{(i,2)}},...,p_{L_{(i,N_\mathrm{par})}},$\\
        $M_{(i,0)},M_{(i,1)},...,M_{(i,N_\mathrm{meta})})$}\;
    }
}

\tcc{Concatenate the output}
$\bm{y} = (~Y_{{0,0}}~...~Y_{{0,N_\mathrm{dat}}}~Y_{{1,0}}~...~Y_{{1,N_\mathrm{dat}}}~...~Y_{{N_\mathrm{set},N_\mathrm{dat}}}~)$\;
\textbf{return} $\bm{y}$\;
}
\textbf{end}
\caption{Master function for global fitting}
\label{alg:master_eq}
\end{algorithm}

\section{Examples}

\subsection{mudpy}

Our first example uses \texttt{mudpy.mdata} to read a \gls{mud} file called \texttt{041200.msr}, located in the current working directory, and corresponding to an \gls{slr} run taken on the \bnmr\ spectrometer in 2019.
\begin{lstlisting}[language=Python]
In [1]: import mudpy as mp                
                                                               
In [2]: data = mp.mdata('041200.msr')

In [3]: data
Out[3]: 
    apparatus: 'BNMR'
         area: 'BNMR'
          das: 'MIDAS'
  description: 16908289
     duration: 622
     end_date: 'Mon Oct 28 01:45:13 2019'
     end_time: 1572252313
          exp: 1424
 experimenter: 'df, ms'
        field: '65500.8(0.0)G'
         hist: mdict: {'B+', 'F+', 'B-', 'F-', ... }
         ivar: mdict: {'BNMR:HVBIAS:POS:RDVOL', 'BNMR:HVBIAS:NEG:RDVOL', ...}
          lab: 'TRIUMF'
       method: 'TD-BNMR'
         mode: '20'
  orientation: ''
          run: 41200
       sample: 'EMIM-Ac (LiCl)'
         sclr: mdict: {'Back%BSegments', 'Front%FSegments', ...}
   start_date: 'Mon Oct 28 01:34:51 2019'
   start_time: 1572251691
  temperature: '243.9(4.1)K'
        title: 'EMIM-Ac (LiCl) , B=65500.8(0.0)G, HV=0 kV, T=200.0(0.7)K (warming), SLR'
         year: 2019
\end{lstlisting}
As shown, the output is nicely formatted, noting that the full output of the \texttt{mdict}s have been truncated to conserve on space in this publication. The object \texttt{mdict} belongs to the \texttt{mudpy.containers} module and inherits from the standard Python dictionary class. \texttt{mdict} allows for dictionary elements to be accessed as attributes. Therefore, \texttt{data.ivar.var\_name} is equivalent to \lstinline[basicstyle=\ttfamily]{data.ivar['var_name']}, subject to restrictions on allowed attribute names (e.g. \texttt{data.hist.F+} is not an allowed syntax; for this special case \texttt{data.hist.Fp} can be used, or for the other helicity state: \texttt{data.hist.Fn}). \texttt{mudpy.containers} also defines the \texttt{mcontainer} base class which is the parent for containers specific to the histogram, variable, scaler, and comment data types. \texttt{mcontainer} has numerous convenience functions such as nice printing: 
\begin{lstlisting}[language=Python]
In [4]: data.ivar['/Sample/read_A']
Out[4]: 
 description: 'SampleA temperature'
        high: 250.383
   id_number: 23
         low: 236.001
        mean: 243.9106082123654
        skew: -831.177521713147
         std: 4.067047463404781
       title: '/Sample/read_A'
       units: 'K'
\end{lstlisting}
as well as redefined mathematical operators that act on the \texttt{mean} or \texttt{data} attribute, as appropriate:
\begin{lstlisting}[language=Python]
In [5]: data.ivar['/Sample/read_A']+10
Out[5]: 253.9106082123654
\end{lstlisting}
The other object of interest defined in \texttt{mudpy.containers} is \texttt{mlist}. This object allows for attribute access to its contents as demonstrated in the following example:
\begin{lstlisting}[language=Python]
In [1]: from mudpy.containers import mlist

In [2]: class example(object):
            def __init__(self, x):
                self.x = x

In [3]: m = mlist([example(i) for i in range(5)])                

In [4]: m.x
Out[4]: array([0, 1, 2, 3, 4])
\end{lstlisting}
If the output is a list of numbers, then it is converted to a \texttt{numpy.ndarray}, otherwise it returns another \texttt{mlist} object.

\subsection{bdata}

The bdata package defines three objects: \texttt{bdata}, \texttt{bjoined}, and \texttt{bmerged}. The first inherits from \texttt{mudpy.mdata} with additional formatting and sorting for variables. Additionally, differing variable names are all mapped to a consistent naming scheme. In the following example, we draw the combined asymmetry (calculated with \Cref{eq:asym_4counter}) for the run we fetched with \texttt{mdata}. One major difference between the two is that \texttt{mdata} requires the full file path to be specified, whereas \texttt{bdata} uses the environment variables \texttt{BNMR\_ARCHIVE} and \texttt{BNQR\_ARCHIVE} to locate the data on the local machine. If the data is not found, then the data is fetched from the archive.\cite{Note2} If the environment variables are not defined, the default location for the data is \texttt{\$HOME/.bdata}. The following short example demonstrates how easily one is able to draw the asymmetry of a \gls{bnmr} run:
\begin{lstlisting}[language=Python]
import bdata as bd
import matplotlib.pyplot as plt

data = bd.bdata(run = 41200, year = 2019)
plt.errorbar(*data.asym('c'))
\end{lstlisting}
The reader is encouraged to read the docstring of \texttt{bdata.bdata.asym} for the many options which vary according to type of data collected.

The \texttt{bjoined} and \texttt{bmerged} objects both take as input lists of \texttt{bdata} objects and combine them. After construction, \texttt{bmerged} behaves exactly like a \texttt{bdata} object, combining both the data and the other variables, allowing for seamless replacement in code written for \texttt{bdata}. This results in a loss of information, namely the individual details of each of the runs. The object \texttt{bjoined} solves this issue, taking advantage of \texttt{mudpy.containers.mlist} to store the information of each run. In this way data from a set of runs is easily combined, however the operation of \texttt{bjoined} is slightly different from \texttt{bdata}, returning lists of the data containers rather than single merged containers. 

\subsection{bfit}

We now show an example of using the fit functions defined in bfit to fit some pulsed beam \gls{slr} data:
\begin{lstlisting}[language=Python]
import bdata as bd                                                                               
from scipy.optimize import curve_fit                                                             
from bfit.fitting.functions import pulsed_strexp

data = bd.bdata(run = 41200, year = 2019)                                                                

# define stretched exponential function
# parameters: 1/T1, beta, amplitude
# (bdata provides common probe lifetimes)
fn = pulsed_strexp(lifetime = bd.life.Li8, 
                   pulse_len = data.get_pulse_s())                    
                           
# calculate combined asymmetry and fit
time, asym, error = data.asym('c')
non_zero = error > 0

par, cov = curve_fit(f = fn, 
                     xdata = time[non_zero], 
                     ydata = asym[non_zero], 
                     sigma = error[non_zero], 
                     absolute_sigma = True) 
\end{lstlisting}

As a simple, if contrived, example of using the global fitter, consider the following:
\begin{lstlisting}[language=Python]
import numpy as np
from numpy.random import randint, random
from bfit.fitting.global_fitter import global_fitter

fn = lambda x, a, b : a * x + b

# make some data, irregularly sized
x = [np.arange(randint(low = 3, high = 10)) for i in range(5)]

# have a varying intercept, shared slope of 2
itrcpt = np.arange(len(x))*5
y = [fn(xval, 2, i) for xval, i in zip(x, itrcpt)]

# random error
dy = [random(len(xval)) for xval in x]

# set up the fitter, sharing the slopes, but 
# different intercepts
gf = global_fitter(x, y, dy, fn, 
                   shared = (True, False))

# do the fit (inputs are passed to curve_fit, 
# after flattening)
gf.fit()

# get the results and draw
par, cov, std = gf.get_par()
global_chi2, chi2 = gf.get_chi()
gf.draw()
\end{lstlisting}
%
 
\section{Summary}

The packages mudpy, bdata, and bfit bring the rapid development style of Python 3 to \musr\ and \bnmr\ analysis, complementing established \musr\ software such as Musrfit. The first package provides an easy interface to \gls{mud} files though the \texttt{mdata} class. The second does the same for \bnmr\ data through the \texttt{bdata} class, but with additional analysis and convenience tools. The third provides both a \gls{gui} and an \gls{api} for \bnmr\ fitting functions and tools such as the global fitter. All three packages are freely available on \gls{pypi} and GitHub.\cite{Note1}
 
\begin{acknowledgments}
The author would like to thank R.~M.~L. McFadden for sharing many of his codes, on which much of the \bnmr\ implementation is based, as well as for many useful and interesting discussions. Additionally, the members of the TRIUMF \bnmr\ group have been instrumental in testing, providing feedback, and suggesting improvements. This work was supported by the QuEST fellowship program.
\end{acknowledgments}



\bibliography{/home/fuji/Documents/Research/Papers/library.bib,references.bib}

\end{document}